\documentclass{aa}  
\usepackage{todonotes}
\usepackage{amssymb}
\usepackage{hyperref}
\usepackage{lineno}
\usepackage{subfigure}
\usepackage{graphicx}
\usepackage{txfonts}
\usepackage{xcolor}

\usepackage[normalem]{ulem}

\usepackage{soul}
\begin{document}    

\title{Anomalous cosmic rays within the inner heliosphere: Observations of helium by the High Energy Telescope onboard Solar Orbiter}
\authorrunning{Xu et al.}
   
   \author{Zigong Xu 
          \inst{1,2}
          \and
          Robert F. Wimmer-Schweingruber \inst{1}
          \and
          Lars Berger \inst{1}
          \and
          Patrick Kühl \inst{1}
          \and
          Alexander Kollhoff \inst{1}
          \and
          Bernd Heber \inst{1}
          \and
          Stephan I. Böttcher \inst{1}
          \and
          Liu Yang \inst{1}
          \and 
          Verena Heidrich-Meisner \inst{1}
          \and
          Roelf Du Toit Strauss\inst{3}
          \and 
          Raúl Gomez-Herrero \inst{4}
          \and
          Javier Rodriguez-Pacheco \inst{4}
          \and
          Daniel Pacheco \inst{5}
          \and
          Richard A. Leske \inst{2}
          }
     \institute{Institute of Experimental and Applied Physics, Christian-Albrechts-University Kiel, 24118 Kiel 
        \and
        California Institute of Technology, Pasadena, CA 91125, USA\\     \email{zgxu@caltech.edu}
        \and
        Centre for Space Research, North-West University, Potchefstroom, 2520, South Africa
        \and 
        University of Alcala, Space Research Group, Alcalá de Henares, Madrid, Spain
        \and 
        Deep Space Exploration Laboratory/School of Earth and Space Sciences, University of Science and Technology of China, Hefei 230026, P.R. China}

   \date{}
  \abstract
   {Radial gradients of cosmic rays are key parameters for understanding the transport of particles in space.
   Solar Orbiter, launched on 2020 February 10, approaches the Sun approximately every half year, with a closest perihelion distance of 0.29 au after the end of 2022 during the nominal mission phase. The two double-ended high energy telescopes (HET) onboard the Solar Orbiter measure energetic particles in the energy range between a few MeV/nuc and a few hundred MeV/nuc, which are dominated by anomalous cosmic rays (ACRs) and galactic cosmic rays (GCRs) during solar quiet times.}     
  {By obtaining the radial gradient of the ACR helium in the inner heliosphere, we advance our understanding of how the transport of the cosmic rays is affected by the particle drift effect and the large-scale magnetic field.}
  {The helium observations at Solar Orbiter/HET between 11.1 and 49 MeV/nuc are analyzed.  Since we focus on quiet time measurements, we remove the periods of solar energetic particle (SEP) events. The intensities are averaged over the Carrington rotation period. The helium observations from the Proton and Helium Instrument (EPHIN) onboard SOHO were utilized as the baseline to correct the long-term variation caused by the solar modulations.}
  {We present the first observation of ACR helium at Solar Orbiter/HET between 2020 February and 2022 July in the inner heliosphere before the sun became fully active. We derive the radial gradient of the ACR helium between 0.3 and 1 au.
  The averaged radial gradient between 11.1 and 49 MeV/nuc is about 22 $\pm$ 4 \% / au and the averaged value between 11.1 and 41.2 MeV/nuc is raised to 32 $\pm$ 8 \% / au after removing the GCR contribution, which is estimated by a GCR model. In addition, the temporal variation of radial gradients indicates that the gradients are increasing with the enhancement of the solar modulation and the increased tilt angle of the heliospheric current sheet.}
  {}
    \titlerunning{ACR helium by Solar Orbiter}
    \maketitle
%
\section{Introduction} \label{intro}

Anomalous cosmic rays (ACRs) are thought to originate as interstellar neutrals, which become ionized in the interplanetary medium, and then convected out to the outer heliosphere by the expanding interplanetary magnetic field, where they are accelerated to become energetic particles.  
\citep[and references therein]{fisk_interpretation_1974, cummings_composition_2002, mccomas_explanation_2006, cummings_composition_2007, giacalone_anomalous_2022,cummings_acceleration_2024}. 

After being accelerated, the energetic particles, i.e., ACRs, re-enter the heliosphere where their motions in the solar wind are determined by their interaction with the IMF. The following physical processes can affect the propagation of particles: (a) diffusion caused by the irregular magnetic field, (b) convection and adiabatic energy loss due to the expanding solar wind, and (c) gradient and curvature drifts in the heliospheric magnetic field. Though these processes have been well described and modeled in the cosmic ray transport equations, and have long been studied for both ACRs and galactic cosmic rays (GCRs) \citep[e.g.,][]{Parker1965Pss,Jokipii1977ApJ, Jokipii1981ApJ, McDonald2001ICRC, Potgieter2013LRSP, giacalone_anomalous_2022,engelbrecht_theory_2022}, the detailed roles of each component when they interact with the varying solar magnetic fields at different scales are still largely unexplored in the inner heliosphere, particularly in the region within 1 au which has hardly been visited by human-made spacecraft.

Among many parameters that spacecraft can measure, the spatial gradients — including both radial and latitudinal components — can specifically provide valuable insights into the variation of the charged particles' drift motion across different solar cycles. As elucidated by theory  \citep[e.g.,][]{Jokipii1977ApJ, Jokipii1979ApJ}, the drift direction of energetic particles reverses when the polarity of the Sun's magnetic field changes, which alternates approximately every 11 years and results in the formation of a 22-year variation \citep{cliver_22year_1996, richardson_22year_1999}. During the positive polarity (hereafter denoted by A+), the positively charged ions drift inward from the south and north pole regions while moving outward along the equatorial plane through the heliospheric current sheet (HCS). Conversely, during the negative polarity (denoted as A-) period, ions drift toward the Sun along the HCS in the equatorial plane and exit through poles of the heliosphere. The drift direction of electrons is naturally opposite to that of ions due to the different charge sign. 
As anticipated by models and calculations, this phase-dependent behavior of cosmic rays might lead to several peculiar characteristics, such as a different radial gradient in ACRs' intensities and an opposite sign of the latitudinal gradient across different cycles \citep[and references therein]{Rankin2022ApJ_oxygen, giacalone_anomalous_2022}

The recent solar activity minimum period between 2018 and 2022 was characterized by an A+ phase. Consequently, the radial gradients in intensities that manifest the drift processes in the heliosphere were expected to be smaller than those in the last solar cycle and those with A- \citep{Potgieter2013LRSP}. However, the measurements from IS$\odot$IS \citep{mccomas_integrated_2016} onboard the Parker Solar Probe (PSP, \cite{Fox_Solar_2016SSRv})
indicate that radial gradients of ACRs oxygen within 1 au between 2018 and 2020 are inconsistent with model predictions and measurements of previous solar minima within the same polarity, but comparable with the gradients of A- phase solar cycles  \citep{Rankin2021ApJ_helium, Rankin2022ApJ_oxygen}, despite the fact that previous observations correspond to the heliospheric region of larger distances, i.e., beyond 1 au. The radial gradients of ACR oxygen determined by Helios \citep{marquardt2018AA} are consistent with PSP's result.
The discrepancies in the radial gradient at different locations reveal the distinct magnetic field environment that may affect the diffusion of charged particles in the inner and outer heliosphere, which warrants further investigations.
In addition, studies have shown unusual behaviors of GCR intensities over the past two solar minima, in which the intensities peaked and reached historically high levels \citep{mewaldt_record-setting_2010, zhao_modulation_2014, Fu2021ApJS}. The plausible explanations include the reduced solar wind dynamic pressure and a less turbulent magnetic field environment \citep{Rankin2022ApJ_oxygen}.

\begin{figure}
    \centering
    \subfigure[]{
     \includegraphics[width=0.75\linewidth]{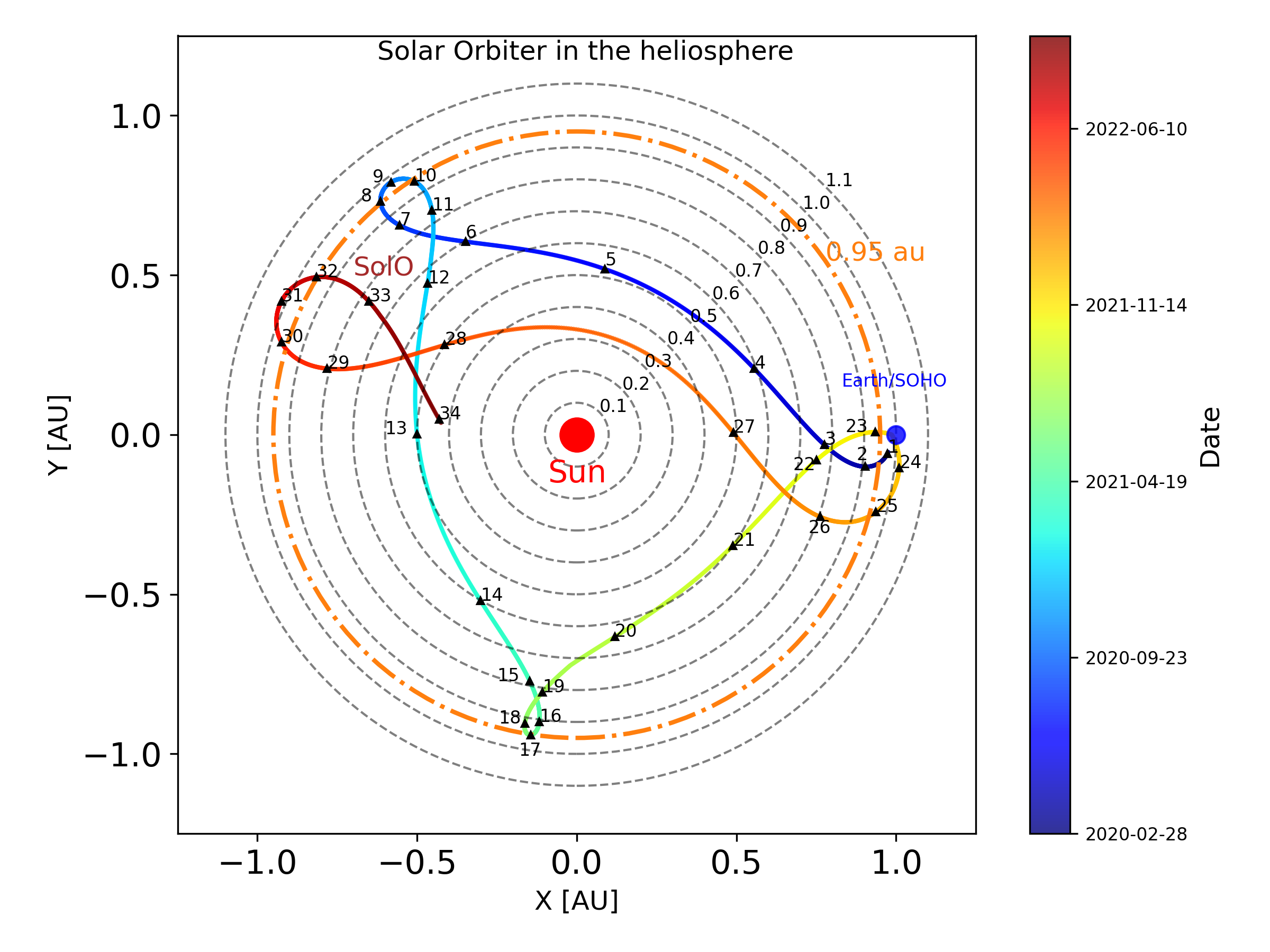}    
     \label{fig:1a}}
     \subfigure[]{
    \includegraphics[width=0.75\linewidth]{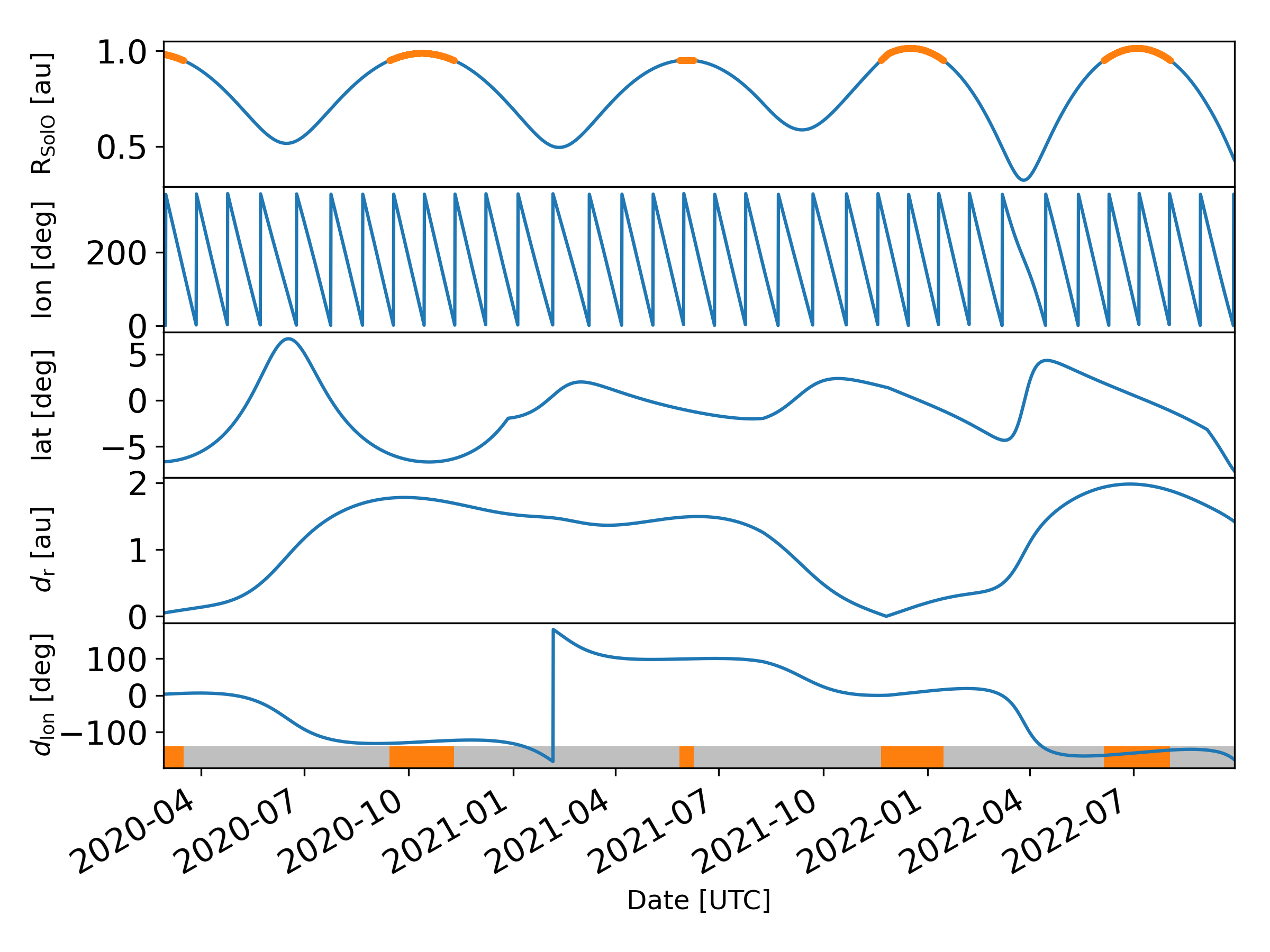}
    \label{fig:1b}
    }
    \caption{(a): The orbit trajectories of Solar Orbiter (2020.02.28 - 2022.08.31) in the Stonyhurst heliographic coordinate system with the Earth fixed at 1 au as the blue dot. The color scale of the twisted trajectories represents the order of time since its launch. The first 4 orbits and 34 Solar Orbiter-perspective Carrington rotations are included in this figure. The orange dashed line represents a circle of 0.95 au radial distance.
   (b): From top to bottom, the panels display the time variation of Solar Orbiter's radial distance,  Carrington longitude and heliographic latitude, together with its relative distance ($d_\mathrm{r}$) and longitudinal ($d_\mathrm{lon}$) separation from Earth. Orange segments over the blue line in the top panel indicate periods when the radial distance of Solar Orbiter exceeded 0.95 au, and likewise for the grey and orange bars at the bottom of the figure.}
    \label{fig:1-Orbit Overview}
\end{figure}

In this study, we aim to deepen our understanding of cosmic ray variations and the above-mentioned discrepancies by utilizing the latest cosmic ray observations from the Solar Orbiter (SolO, \cite{Mueller-2020-SolO}), which was launched on 2020 February 10 in the middle of this solar activity minimum period and travels between 0.28 and 1 au. We focus our studies on the time period between 2020 and 2022 during the second half of the solar activity minimum between solar cycles 24 and 25. 
In Fig.~\ref{fig:1-Orbit Overview}, we show the orbit information of Solar Orbiter during the first 4 orbits from February 2020 to September 2022 before the solar activity maximum. The top panel (a) depicts the orbit track of Solar Orbiter in the Heliographic-Earth-Equatorial (HEEQ)
coordinate system. In this coordinate system, the Sun, shown as a red dot, is positioned in the center while the Earth (blue dot) is fixed at 1 au. The trajectory of Solar Orbiter is twisted in this coordinate system. The color bar indicates the time order, and the numbers are the rotations of the Sun from Solar Orbiter's perspective i.e., Solar Orbiter-perspective Carrington rotation, since it launched, which are marked at the beginning of each rotation and labeled next to the orbit track. Therefore, the flow direction of Solar Orbiter can be inferred from the increment in number and the change of colors. The orange dot-dashed circle line has a radius of 0.95 au. The five sub-panels in Figure~\ref{fig:1b} display the temporal variation of the radial distance of Solar Orbiter from the Sun, Carrington longitude, heliographic latitude of Solar Orbiter, and the distance as well as the longitudinal separation between Solar Orbiter and Earth. It should be noted that during this period, the latitude of Solar Orbiter was still constrained within a range of $\sim$ $\pm$8 degrees of the ecliptic plane. After February 2025, Solar Orbiter has gradually been moving out of the ecliptic plane and has been observing from a higher latitude region.
The orange portions in the first panel of Figure~\ref{fig:1b} indicate when Solar Orbiter's radial distance was greater than 0.95 au. Data from these time slots will be used to conduct a cross-comparison of energetic particle spectra between Solar Orbiter and various scientific instruments located near Earth.

Solar Orbiter's unique measurements of cosmic rays within 1 au, together with those from PSP, provide a valuable opportunity to investigate the unknown turbulent interplanetary environment that affects the propagation of cosmic rays close to the Sun. By deriving the radial gradient of energetic particles, these combined observations open a new regime for studying particle transport processes \citep{giacalone_anomalous_2022}. 
This paper is organized as follows. Section~\ref{sec:instrument} introduces the instrument and data we used in this study. In Section~\ref{sec:obs}, we first present an overview of the first 2.5 years of the ACR helium\footnote{This always refers to helium-4 in this paper.} observations, after excluding the sporadic solar energetic particle (SEP) events that occurred in this period, and provide a cross-comparison among different instruments. Subsequently, we provide the critical results of ACR helium radial gradients at different energies. We conclude the paper with a summary and discussion section.

\section{Instruments}\label{sec:instrument}

This study is based on the measurements of energetic particles from the high-energy telescope (HET), which is the higher energy part of the Energetic Particle Detector (EPD) suite onboard Solar Orbiter\footnote{Source: \href{https://soar.esac.esa.int/soar/}{https://soar.esac.esa.int/soar/}} \citep{RodriguezPacheco-2019-EPD}. HET consists of two double-ended telescopes measuring charged particles in the energy range from tens of MeV/nuc to a few hundred MeV/nuc. Within this energy coverage, the charged particles are dominated by ACRs and GCRs for quiet time, rather than energetic particles that originate from the Sun. 
Two identical HETs are mounted perpendicularly in the spacecraft's ecliptic and polar plane, measuring particles coming from four directions.

Similarly to the results deriving from the suprathermal ion spectrograph (SIS), which have relatively larger uncertainties in radial gradient \citep{Mason-2021-SolOQuietTime}, the accuracy of the radial gradients of helium calculated from HET is also limited by the counting statistics in each energy channel. As a result, we employ the following two methods to reduce uncertainties. Firstly, we average measurements from four HET apertures, based on the assumption that cosmic rays are nearly isotropically distributed within those directions. By doing so, we can reduce the statistical uncertainty by a factor of 2 since we increase the count rate by a factor of 4. Then, we re-bin the fine energy channels of HET into four new energy channels spanning the range from 10 to 50 MeV/nuc. The new energy bins are 11.1 - 19.4 MeV/nuc, 19.4 - 29.5 MeV/nuc, 29.5 - 41.2 MeV/nuc, 41.2 - 49.0 MeV/nuc.

In addition to Solar Orbiter/HET, we employ the helium measurements from the new L3 data product\footnote{Source: \href{http://ulysses.physik.uni-kiel.de/costep/level3/l3i/}{http://ulysses.physik.uni-kiel.de/costep/level3/l3i/}}\citep{kuhl_revising_2019} of Electron Proton Helium Instrument (EPHIN) onboard the Solar and Heliospheric Observatory \citep[SOHO,][]{muller-mellin_costep_1995} and the level 2 data\footnote{Source: \href{https://izw1.caltech.edu/ACE/ASC/level2/lvl2DATA_SIS.html}{https://izw1.caltech.edu/ACE/ASC/level2/lvl2DATA\_SIS.html}} of Solar Isotope Spectrometer \citep[SIS,][]{stone_1998_ace} onboard the Advanced Composition Explorer (ACE) at the first Lagrange point (L1), as well as those from the Lunar Lander Neutron and Dosimetry (LND)\footnote{Source: \href{https://www.ieap.uni-kiel.de/et/change4/}{https://www.ieap.uni-kiel.de/et/change4/}} experiment \citep{Wimmer2020SSRv} onboard the Chang'E-4 lander on the lunar far-side surface, for spectra comparison. In the end, only helium observations from SOHO/EPHIN serve as the baseline of the long-term variation due to the solar modulation and are used to calculate the radial gradient within 1 au, as it has a complete energy coverage between 10 and 50 MeV/nuc, within which we conduct our analysis for both Solar Orbiter and SOHO/EPHIN. See Section~\ref{sec:radialgradient} for details of the analysis.

\section{Observation}
\label{sec:obs}

The observations are described in this section, including the cross-comparison for ACR helium spectra within multiple instruments in Sec.~\ref{sec:obs-1}, the temporal variations of ACR helium in Sec.~\ref{sec:obs-2} and calculation of radial gradients in Sec.~\ref{sec:radialgradient}.

\subsection{ACR helium near 1 au}
\label{sec:obs-1}

\begin{table*}[!htb]
\caption{Periods when Solar Orbiter was positioned between 0.95 and 1 au.}
    \centering
    \begin{tabular}{|c|c|c|c|c|}
    \hline
    No. & Start time & End time & Distance to Earth  (au) & ratio(HET/EPHIN) \\
    \hline
    1   & 2020-02-28 & 2020-03-16   & 0.07  & 1.10 $\pm$ 0.05\\
    \hline
    2   & 2020-09-14 & 2020-11-10   & 1.76  & 1.10 $\pm$ 0.03\\
    \hline
    3   & 2021-05-27 & 2021-06-09   & 1.49  & 0.90 $\pm$ 0.06\\
    \hline
    4   & 2021-11-21 & 2022-01-15   & 0.13  & 1.08 $\pm$ 0.04\\
    \hline
    5   & 2022-06-05 & 2022-08-02   & 1.96  & 1.0  $\pm$ 0.06 \\
    \hline

    \end{tabular}
 
 \tablefoot{The radial distances between Solar Orbiter and Earth are listed in the fourth column. SEP events and ICME periods were excluded from the spectra comparison. The last column gives the averaged intensity ratio between HET and EPHIN.}
   \label{tab:1AU_period}
\end{table*}

\begin{figure}
    \centering
    \resizebox{\hsize}{!}{\includegraphics{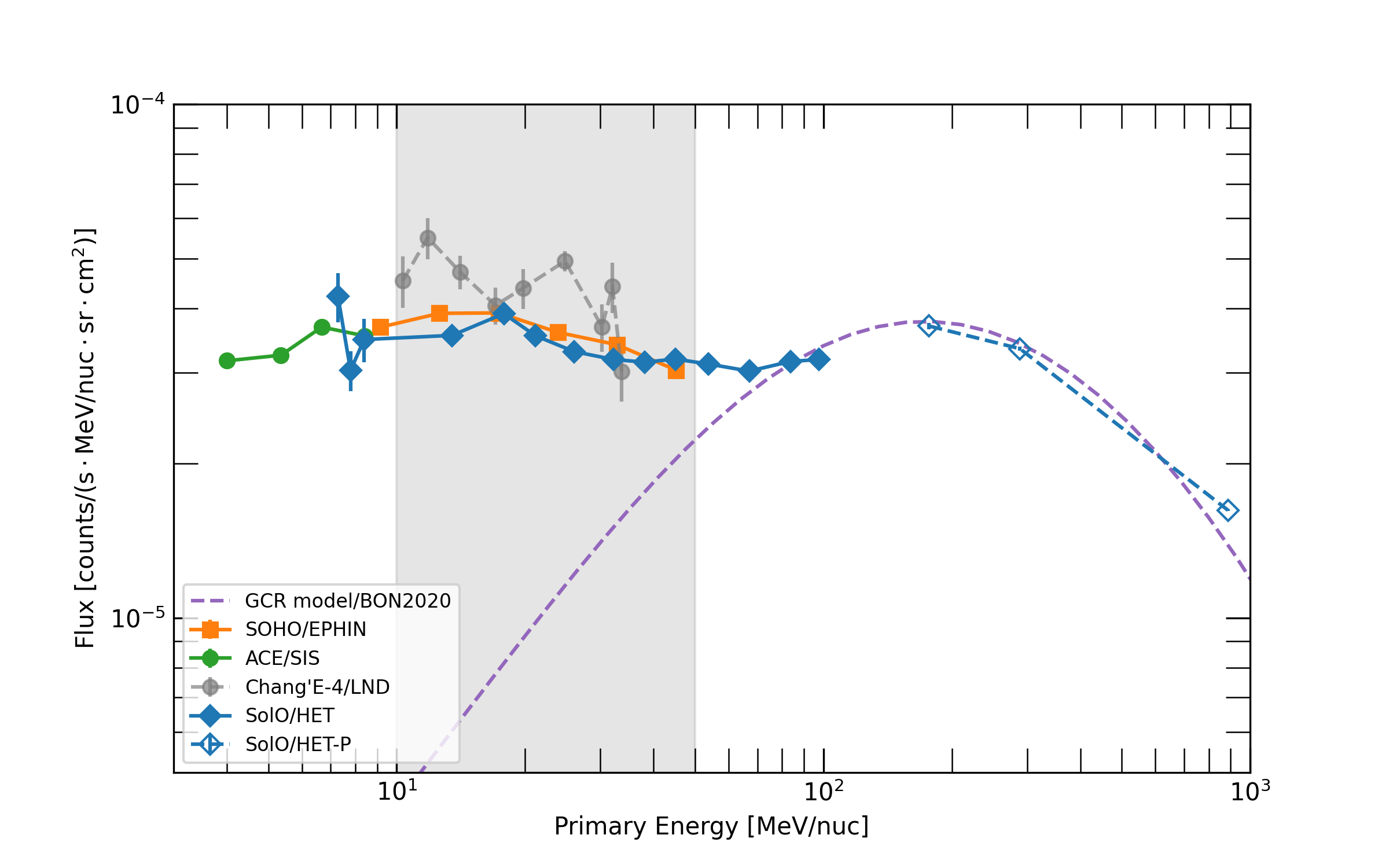}}

    \caption{The quiet time helium spectra of Solar Orbiter/HET for stopping (filled blue diamonds) and penetrating (empty blue diamonds) particles, of ACE/SIS (green circles), and SOHO/EPHIN (orange squares) at L1, as well as Chang'E-4/LND (grey circles) measured at the lunar surface. The helium spectra were averaged over periods when Solar Orbiter was located between 0.95 and 1 au, as given in Tab.~\ref{tab:1AU_period}. The GCR spectrum predicted from the BON2020 model \citep{slaba_badhwaroneill_2020} is given as the dashed line. Error bars indicate the statistical uncertainties.}
    \label{fig:2-Helium-Spectra-1au}
\end{figure}

Before extracting the subtle radial variations of ACR helium from measurements of Solar Orbiter, it is crucial to first cross-compare the observations among different instruments, particularly during periods when Solar Orbiter was close to them.
In Fig.~\ref{fig:2-Helium-Spectra-1au}, we display ACR helium spectra from SOHO/EPHIN (orange squares, 7.8 - 53 MeV/nuc ), ACE/SIS (green circles, 3.4 - 7.9 MeV/nuc), Chang'E-4/LND (grey circles, 9.7 - 34.5 MeV/nuc) and most importantly the spectrum from Solar Orbiter (blue diamonds) which has a broader energy coverage from a few MeV/nuc to a few hundreds MeV/nuc. The filled diamonds represent stopping helium particles with energies between 7.2 and 104 MeV/nuc, and the empty diamonds are derived from penetrating channels where helium with energies between 147 MeV and 2 GeV are measured. This energy range is dominated by GCRs rather than ACRs; hence, open diamonds are used. When we analyze ACRs in this paper, we only focus on the energy range between 10 and 50 MeV/nuc where both SOHO/EPHIN and Solar Orbiter/HET have measurements, which is indicated as the gray-shaded region in Fig.~
\ref{fig:2-Helium-Spectra-1au}.

The spectra in Fig.~\ref{fig:2-Helium-Spectra-1au} are averaged over periods when Solar Orbiter was located between 0.95 au and 1 au and exclude the contamination of SEPs to the background intensity, as discussed in Sec.~\ref{sec:obs-2} and Appendix. We identify five time periods from 2020 to 2023, with their specific start and end times listed in Tab.~\ref{tab:1AU_period}. In addition, the fourth column lists the radial distance separation between Solar Orbiter and Earth. 
The spectrum in gray circles in particular represents the first ACRs helium observations on the lunar far-side surface by Chang'E-4/LND. 
The past studies have confirmed that there was no significant discrepancy between the high-energy particle environment on the lunar surface and the deep space near the L1 point, during both SEP events and solar quiet time \citep{xu_first_2020, Xu_2022FrASS}, which is expected since the Moon lacks a dense atmosphere and a global magnetic field.

The quiet time spectra from SOHO/EPHIN, ACE/SIS and Chang'E-4/LND, all measuring particles in the near Earth region, show a general consistency across different energies during those specific time intervals. The averaged intensity ratio of Solar Orbiter/HET to the SOHO/EPHIN is 0.962 $\pm$ 0.016 within the energy range between $\sim$10 and $\sim$50 MeV/nuc. The spectrum of LND given in gray circles seems higher than the orange one from SOHO/EPHIN at the same energy range. One possible contributor of the enhanced intensity may be associated with the backsplash helium generated by interactions of GCRs with the lunar soil; however, this interpretation remains tentative and requires further validation, which is out of the scope of this study. As a result, LND data were exclusively employed here for overall spectral comparison and excluded from the following calculation of the ACR helium radial gradient.

The spectra comparison among four instruments shown in Fig.~\ref{fig:2-Helium-Spectra-1au} shows that the helium spectrum that is averaged from four Solar Orbiter/HET apertures agrees with the counterparts at SOHO/EPHIN and ACE/SIS. Meanwhile, the higher energy spectra at HET show a good agreement with the prediction of the GCR spectrum by the Badhwar‐O'Neill (BON-2020) model \citep{slaba_badhwaroneill_2020}. The model spectrum indicates the averaged intensities during periods listed in Tab.~\ref{tab:1AU_period} and is shown as a dashed line.

\begin{figure*}
    \sidecaption
    \includegraphics[width=12cm]{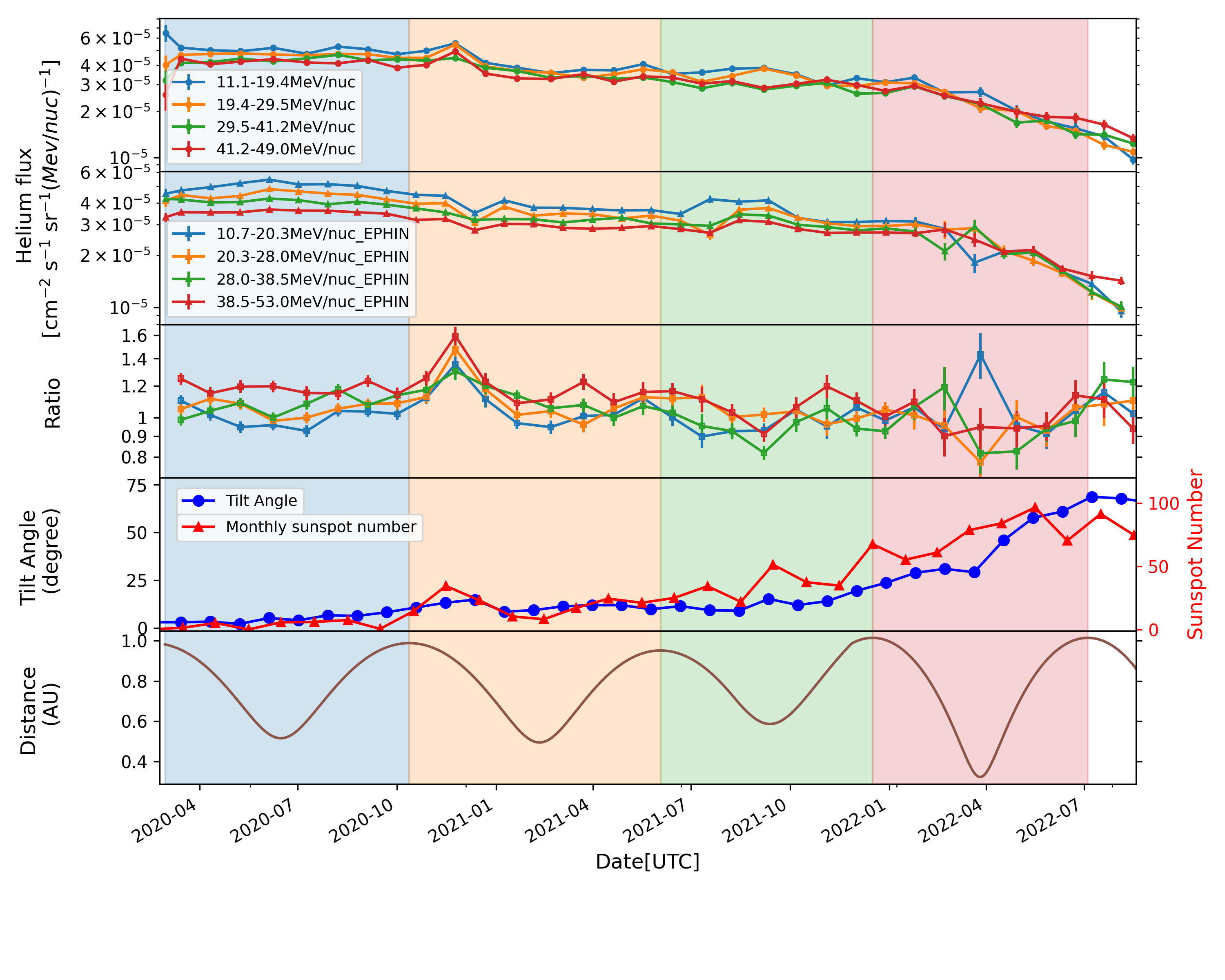}
    \caption{From top to bottom: the averaged helium intensity between $\sim$ 10 and $\sim$ 50 MeV/nuc measured by Solar Orbiter/HET (top panel) and SOHO/EPHIN (second panel) over the corresponding Carrington rotations between 2020 February and 2022 September; the intensity ratio of Solar Orbiter/HET to SOHO/EPHIN (third panel); the monthly averaged tilt angle of HCS and sunspot number (fourth panel); Solar Orbiter's radial distance (bottom panel).}
    \label{fig:overview_helium_intensity}
\end{figure*}

\subsection{Temporal variation of the ACR helium intensity between 2020 and 2022}
\label{sec:obs-2}

To obtain ACR helium, we consider the following two major factors that could affect the variation of the helium intensity.
The first one is SEPs that are generated from the magnetic reconnection process during the solar flare eruptions and from the coronal mass ejection (CME) driven shocks via two different acceleration mechanisms \citep{desai_large_2016}. The solar-originated helium could significantly enhance the helium intensity level and is the foremost disturbance to deal with in our study. 
Luckily, this study focuses on observations before September 2022, a period when solar activity remained low and only a few SEP events influenced helium intensity level. After mid-2022, many extremely intense eruptions, together with large SEP events, occurred and were observed by multiple spacecraft. A list of SEP periods is given in the appendix. The method, as well as the data we used to determine the SEP periods, are also described.  We remove all those SEP periods in our analysis and the helium spectra in Fig.~\ref{fig:2-Helium-Spectra-1au} are the results of quiet time without SEP events and agree with our expectation of the quiet time spectra.

The second is the short-term but periodical modulation caused by the (recurring) compressed regions induced by the fast-slow solar wind stream pair emitted from the solar corona, which is known as stream interaction regions (SIRs) or corotating interaction regions (CIRs) \citep{Burlaga1974JGR, Gosling1976JGR, Richardson2004SSRv}.
These structures occurred periodically within one solar rotation and changed the solar wind pressure, density, level of magnetic turbulence and, as a result, changes the diffusion and scattering level of particles. Those changes in local conditions modify the transport of cosmic rays, resulting in a decrease in intensity \citep{Richardson2004SSRv, Richardson-2018}, even though such variations are not as significant as SEPs, and long-term solar modulations. It should be noted that such effects could even extend to the high latitude, according to the previous observations \citep{McKibben_balogh_modulation_1999}, which will be considered in future works after Solar Orbiter moves out of the ecliptic plane.
The periodical modulation of such solar structures on the cosmic-ray intensity can be reduced by averaging particle fluxes over a Carrington rotation presumably, as suggested by \cite{Rankin2021ApJ_helium, Rankin2022ApJ_oxygen}.
Naturally, the Carrington rotation at Earth represents a full rotation of the Sun of a standardized period of 27.3 days. Such period at Solar Orbiter is different than that at Earth, due to Solar Orbiter's tangential velocity. From Solar Orbiter's point of view, one full solar rotation is defined as the period between two adjacent peaks of Solar Orbiter Carrington longitude, which is depicted in the second panel from the top of Fig.~\ref{fig:1-Orbit Overview}. The Solar Orbiter-perspective Carrington rotation periods range between 26.6 and 35.8 days, though the characteristic of the recovery period vary with energy and species. We note that the potential differential modulation of cosmic-ray intensities by CIRs/SIRs at different heliocentric locations within 1 au is not accounted for in the present analysis \citep{luo_numerical_2020, luo_numerical_2024}.

It is worth noting that the interplanetary coronal mass ejections (ICMEs) observed at Solar Orbiter and SOHO are excluded from the respective datasets separately, due to their potential modulation on the cosmic ray intensity \citep[e.g.,][]{richardson_galactic_2011, dumbovic_cosmic_2011, janvier_two-step_2021}. We identify the ICMEs that pass by the Solar Orbiter and the Earth separately from the existing ICME list provided by HELIO4CAST inter-planetary coronal mass ejection catalog \citep[ICMECAT,][]{mostl_prediction_2020, Moestl2020}. In addition, we exclude an additional 1.5 days of post-ICME intervals, as a simplifying approximation of the Forbush decrease recovery time, although the characteristics of the recover phase could vary with particle energy and species and not all ICMEs show a measurable modulation in the ACR intensity.

After removing SEP periods and ICME periods, as well as averaging the remaining EPHIN and HET helium intensity profile over the corresponding Carrington rotation at Earth and Solar Orbiter independently, as suggested by \cite{Rankin2021ApJ_helium}, we obtain the top two panels of Fig.~\ref{fig:overview_helium_intensity}, displaying the Carrington-averaged helium flux profile between 10 and 50 MeV/nuc for Solar Orbiter/HET and SOHO/EPHIN in four energy bins.

A noteworthy trend is that averaged fluxes decrease over time, due to the enhanced solar modulation during the early phase of the new solar cycle. 
In the fourth panel of Fig.~\ref{fig:overview_helium_intensity} we present the temporal variations of the monthly sunspot number\footnote{Source: WDC-SILSO, Royal Observatory of Belgium, Brussels, DOI: https://doi.org/10.24414/qnza-ac80} \citep{SILSO_Sunspot_Number} and the averaged tilt angle of the heliospheric current sheet (HCS) calculated from a model using a radial boundary condition at the photosphere \footnote{Source: http://wso.stanford.edu/Tilts.html}, which indicates the maximum extent in latitude that could be reached by the computed HCS. These parameters show how the solar modulation changed significantly from the beginning to the end of our analysis period. The averaged monthly sunspot number reached about 100 in the middle of 2022, and similarly, the tilt angle concurrently increased to more than $70^{\circ}$.
Consequently, by the end of this period the helium intensities drop to roughly one-fifth of its peak level observed at the start of the mission and the ACR intensity further continues to decrease over the following years until the current solar activity maximum.

To isolate the intensity variations associated with the changing radial distance of Solar Orbiter, it is necessary to remove long-term trends from the measurements. It is reasonable to assume that the slow solar modulation affects the measurements within 1 au in a similar same way, therefore its influence can be canceled out in the intensity ratio.
We use the EPHIN-measured intensities as a reference baseline and calculate the HET-to-EPHIN intensity ratios for four individual energy channels.
We linearly extrapolate the Carrington-averaged intensity of EPHIN from the center of the rotation to the central time of Solar Orbiter-perspective Carrington rotation, to account for the time offset within two observations.
The resulting ratios, shown in the third panel of Fig.~\ref{fig:overview_helium_intensity}, display systematic variability as a function of radial distance and will be used to derive the radial gradients in the following section.
The bottom panel shows the variations of Solar Orbiter's radial distance as a function of time, to facilitate a direct comparison between the intensity ratio which exhibits clear periodic pattern, and the spacecraft's radial distance.
The different colored blocks separate four successive orbits of Solar Orbiter.

\begin{figure*}[!htb]
    \sidecaption
    \includegraphics[width = 12cm]{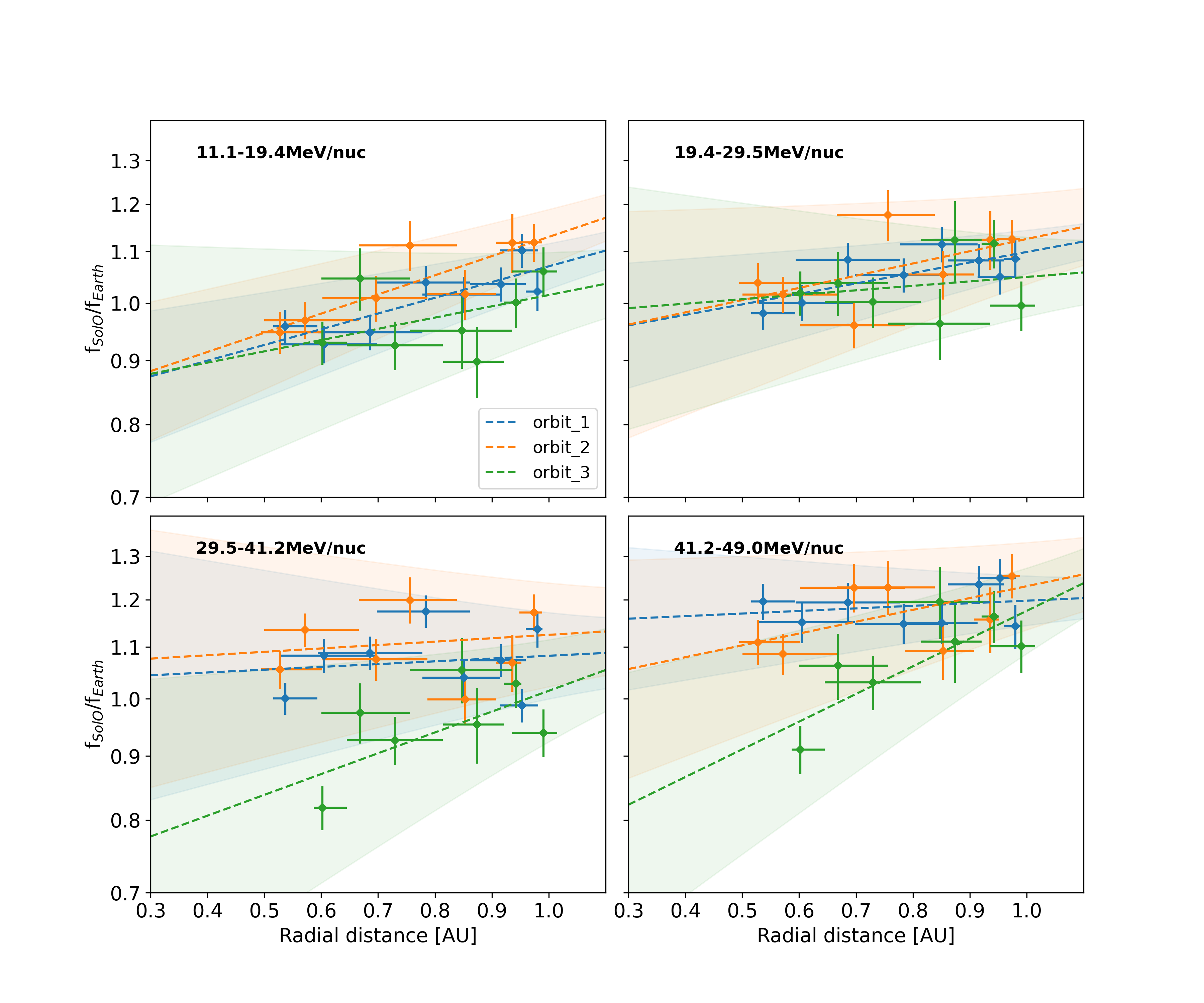}
    \caption{Helium intensity ratios between Solar Orbiter/HET and SOHO/EPHIN as a function of Solar Orbiter’s radial distance during the first three orbits. The dashed lines are the fitted lines for each orbit, which are indicated by colors. The y-axis is shown on a logarithmic scale.} 
    \label{fig4:ratio_radialgradient}
\end{figure*}

\subsection{Radial gradient}\label{sec:radialgradient}

Based on the method from \citep{Vos2016SoPh..291.2181V} and \citep{Rankin2021ApJ_helium} , the radial gradient of cosmic rays can be expressed as:
\begin{equation}
    \mathrm{g_r = \frac{1}{f}\frac{\partial{f}}{\partial{r}} = \frac{\partial{ln} f}{\partial{r}}} \enspace ,
    \label{eq:radial_gradient}
\end{equation}
where $\mathrm{g_r}$ denotes the differential radial gradient component of intensity, i.e., the change of the differential flux ($\mathrm{f}$), with respect to radial distance ($\mathrm{r}$).
$\mathrm{f}$ is the de-trended flux after considering the long-term solar modulation. 
In practice, we then fit the following function to the measurements using the least squares method,
\begin{equation}
    \mathrm{ln\frac{f_{SolO}}{f_{SOHO}} = g_r\cdot (r - 1)}
    \label{eq:fitted}
\end{equation}
where $\mathrm{f_{SolO}}$ and $\mathrm{f_{SOHO}}$ denote the intensity averaged over a Carrington rotation as measured by Solar Orbiter and SOHO, respectively. $\mathrm{r}$ is the radial distance of Solar Orbiter. We assume that the radial gradient of ACR helium is approximately constant within 1 au between Earth and Solar Orbiter.

Fig.~\ref{fig4:ratio_radialgradient} shows the scatter plots of the intensity ratios against radial distance for the first three orbits in separate colors,the same used for the different orbits in Fig.~\ref{fig:overview_helium_intensity}.
The linearly fitted dashed lines at different energies for the first three orbits, together with their 95 \% confidence intervals are depicted accordingly in the same colors, respectively.
When the y-axis is given in the log scale, the fitted lines are straight. 
It should be noted that measurements from the fourth orbit were excluded from the average radial gradient calculations, as their results carry extremely large uncertainties due to the limited counting statistics and are therefore not considered reliable. Outlier data points with ratio values significantly deviating from the mean values are excluded prior to performing the fit. Only two data points are removed. One is the peak ratio point in the second orbit that occurs near November 2020, and the other is the abnormal data point in the forth orbit near April 2022. The former might be due to the differential modulations at Solar Orbiter and Earth, as we discuss later, while the latter is because of the removal of measurements due to a number of SEP events within that Carrington rotation. Only a day or two measurements remain and cannot reflect the ACR helium intensity level of the whole Carrington rotation.
In Tab.~\ref{Tab:radialgradient_1}, we summarize the fitting results for the four orbits between 2020 February and 2022 July, as well as the average radial gradient derived for the first three orbits.

\begin{figure}
    \resizebox{\hsize}{!}{\includegraphics{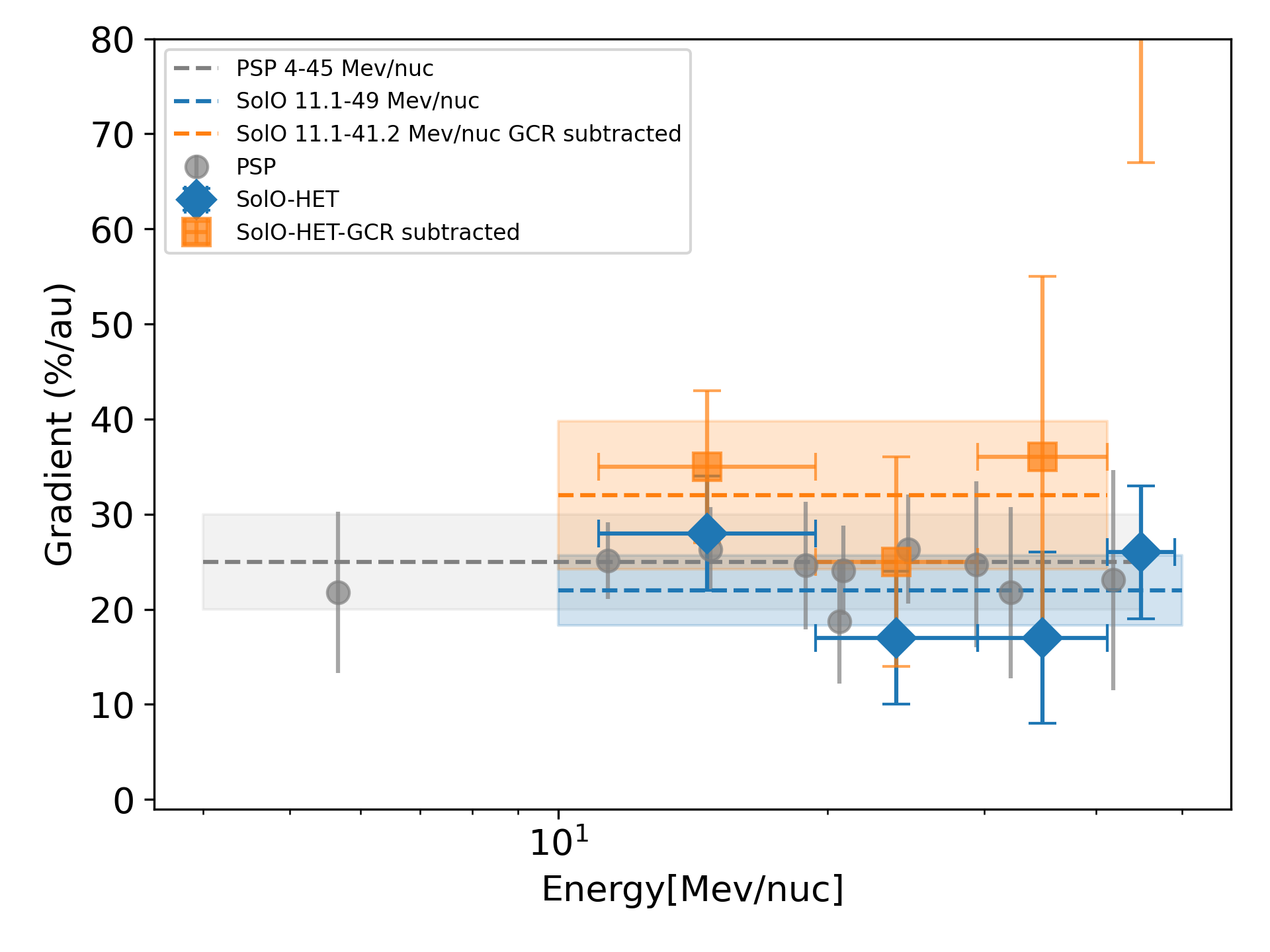}}
    \caption[Energy dependency of the helium radial gradient]{The radial gradients of ACR helium are shown as a function of energy between 11.1 and 49 MeV/nuc. The blue points are the fitted results as in Fig.~\ref{fig4:ratio_radialgradient} and the orange squares further exclude the GCR contamination which are estimated from the BON2020 model, accompanied by their error bars. 
    The grey circles represent the radial gradients obtained from PSP \citep{Rankin2021ApJ_helium}. The dashed lines are the averaged gradient over the corresponding energy range. }
    \label{fig5:gradient_energy}
\end{figure}

Fig.~\ref{fig5:gradient_energy} displays the energy dependence of the ACR helium radial gradients.
The results for Solar Orbiter averaged over the first three orbits are shown as blue diamonds.  We report radial gradients of 28$\pm$ 6 \%/au, 17$\pm$ 7 \%/au, 17$\pm$ 9 \%/au and 26 $\pm$ 7 \%/au for helium energies of 11.1 - 19.4 MeV/nuc, 19.4 - 29.5 MeV/nuc, 29.5 - 41.2 MeV/nuc and 41.2 - 49 MeV/nuc respectively, with an overall average of 22 $\pm$ 4 \% /au between 11 - 49 MeV/nuc, which is shown as the blue horizontal line together with the uncertainties shown as blue shaded region.
Such gradients are consistent with the measurements at PSP within the uncertainties \citep{Rankin2021ApJ_helium}, which are given as the grey circles. PSP results are derived from measurements between 2018 and 2019 in the same solar activity minimum but at an earlier phase. The averaged radial gradient of PSP measured ACR helium in the energy range of 4 - 45 MeV/nuc is about 25$\pm$5 \%/au. 
We also consider the impact of GCR contribution on the ACR radial gradients. The results are shown as orange squares, which we will discuss in the next section.

\begin{table*}[!htb]
    \centering
    \caption{The radial gradients of ACR helium (\%/au) between 0.5 and 1 au during the solar activity minimum 24/25.}
    \begin{tabular}{|c|c|c|c|c|c|}
    \hline
    Number of Orbit     & 1               & 2              & 3               & 4  & average (1 - 3) \\
    \hline
                        &20200228 -      & 20201012-        & 20210602-    &  20211216-   &\\  
    Energy (MeV/nuc)    & 20201012        &  20210602       & 20211216      &  20220704  & \\
    \hline
    11.1 - 19.4 &  29 $\pm$ 8 & 35 $\pm$ 8 & 21 $\pm$ 15 & 10 $\pm$ 27  & 28 $\pm$ 6\\
    \hline
    19.4 - 29.5 &  19 $\pm$ 7 & 22 $\pm$ 13 & 8 $\pm$ 14 & 21 $\pm$ 21  & 17 $\pm$ 7\\
    \hline
    29.5 - 41.2 &  5 $\pm$ 15 & 6 $\pm$ 15 & 38 $\pm$ 19 & 23 $\pm$ 37 & 17 $\pm$ 9\\
    \hline
    41.2 - 49.0 &  5 $\pm$ 8 & 22 $\pm$ 13 & 51 $\pm$ 16 & 26 $\pm$ 31 & 26 $\pm$ 7\\
    \hline
    \end{tabular}

    \label{Tab:radialgradient_1}
\end{table*}

\section{Summary and discussion}

By removing the disturbance of SEPs, correcting the long-term solar modulation, and averaging the intensity over the Carrington rotations to mitigate the impacts of CIR and SIR, we obtain the ACR helium measured by Solar Orbiter/HET and derive the radial gradient for ACR helium in the inner heliosphere by fitting Eq.~\ref{eq:fitted}. 
The helium spectra obtained by Solar Orbiter/HET, SOHO/EPHIN, and ACE/SIS, are generally consistent with each other when Solar Orbiter was close to 1 au.
The mean radial gradient of 22 $\pm$ 4 \% /au for the energy range of 11 - 49 MeV/nuc is consistent with PSP, which reported a gradient of 25 $\pm$ 5\%/au over energies of $\sim$ 4 to $\sim$ 45 MeV/nuc during the same solar minimum, considering the uncertainties. This agreement also holds for the energy-dependent radial gradients.

Although a general consistency has been established, it is still worth emphasizing the observational discrepancies between Solar Orbiter and PSP, which manifest as the differences in observation time and spatial coverage in the heliosphere. 
This study focused on the observations from early 2020 to mid-2022, while PSP results \citep{Rankin2021ApJ_helium} were derived from the PSP observations during its first three orbits between 2018 August and 2019 November, a period right before the Solar Orbiter mission. There is no temporal overlap between the two studies, despite the fact that both Solar Orbiter and PSP were in the same solar minimum period. As the solar cycle evolves, the solar modulation, and the resulting radial gradient are not necessarily the same. As shown in the second-to-bottom panel of Fig.~\ref{fig:overview_helium_intensity}, the variation of the sunspot number and the tilt angle of the HCS, which act as a proxy of the solar modulation, has vastly changed from the first to the fourth orbit of Solar Orbiter, and the strength of solar modulation reached a very high level at the end of our observations. 
In addition, as shown in Fig.~\ref{fig4:ratio_radialgradient}, all data points used in our fitting analysis lie beyond 0.5 au. In contrast, the PSP observations cover a wider radial range, from 1 au down to 0.16 au. This discrepancy may partially explain the relatively larger error bars for each individual orbit (see Tab.~\ref{Tab:radialgradient_1}), compared with those derived from PSP data.

In addition to ACR helium, the GCR helium is also non-negligible in the energy range from a few tens to a few hundreds MeV/nuc. GCR helium intensities increase with energies and typically peak at a few hundred MeV/nuc. 
Therefore, the impact of the GCR helium on the radial gradient should also be considered within the 11.1 - 49 MeV/nuc energy range, particularly toward the higher energy end.  
Meanwhile, the turbulence in the interplanetary medium that varies with solar activities may affect the intensity variation of ACRs and GCRs differently. For instance, when turbulence condition was reduced during the 2019 solar minimum, the intensity of GCRs peaked at higher values than in earlier solar minima while ACRs did not exhibit corresponding enhancement \citep{strauss_modulation_2023}. Therefore, it is essential to properly subtract GCR intensities from the measured ACR intensities in order to derive the accurate radial gradients. 
\cite{Rankin2021ApJ_helium} found that the radial gradient increases after subtracting the assumed GCR intensity. Hence, they concluded that the reported ACR radial gradient represents only a lower limit. This conclusion also holds for this study. To address this, we removed the same estimated GCR intensities from both Solar Orbiter and SOHO measurements.
The GCR intensities are calculated from the Badhwar-O'Neill 2020 (BON2020) GCR Model \citep{slaba_badhwaroneill_2020} based on the varying solar modulations. We assume the same intensities at both locations at a given time by simply ignoring the minimal radial gradient of GCRs within 1 au \citep{marquardt2019, rankin_galactic_2022}.
Although different GCR models may yield slightly different predictions, they generally predict similar intensities during solar minimum conditions, which are comparable to measurements \citep{Xu_2022FrASS, liu_comprehensive_2024}. Hence, the selection of GCR models is not expected to significantly influence our results. 
Note that the purple dashed line shown in Fig.~\ref{fig:2-Helium-Spectra-1au} is the averaged model spectrum over the same interval as the measurements, which closely reproduces the Solar Orbiter GCR helium intensities at penetrating-channel energies above several hundred MeV/nuc. Such an agreement further confirms the reliability of the model spectrum during these periods.

We derive the corrected radial gradients at different energies which are given as orange data points in Fig.~\ref{fig5:gradient_energy}. The radial gradients at all four energies become higher. The new radial gradients are 35 $\pm$ 8 \% /au, 25 $\pm$ 11 \% /au, 36 $\pm$ 19 \%/au, and 90 $\pm$ 23 \%/au for the energy range 11.1- 19.4 MeV/nuc, 19.4 - 29.5 MeV/nuc, 29.5 - 41.2 MeV/nuc and 41.2- 49.0 MeV/nuc respectively.
The significantly enhanced radial gradient at the highest energy bin is not reliable considering the existence of a larger portion of GCR particles whose intensity depends on the adopted GCR model. Deriving ACR helium and its radial gradients within this energy range is challenging and subject to a large uncertainty. Therefore, we simply ignore this outlier data point from the calculation of the average gradient. Future observations with longer accumulation times and improved GCR models might help refine these uncertainties.

Consequently, we obtain an average gradient of 32 $\pm$ 8 \%/au for 11.1 - 41.2 MeV/nuc, which is plotted as orange dashed line and the corresponding shaded region in Fig.~\ref{fig5:gradient_energy}. Unsurprisingly this value is consistent with that derived from PSP measurements within the uncertainty, which reported a gradient of 34.3 $\pm$ 5.6\%/au for 4.0 - 32.0 MeV/nuc derived from PSP/LET and 44.7 $\pm$10.2 \%/au for 13.4-45.3 MeV/nuc from PSP/HET, after subtracting the GCR contributions.

The interplay between particles and the large-scale magnetic field governs the transport of the cosmic rays in space, leading to the varying latitudinal and radial gradients of the cosmic rays \citep{cummings_latitudinal_1987,cummings_anomalous_1995, cummings_anomalous_1999,stone_tilt_1999, cummings_radial_2009}. The magnetic field structures in the inner heliosphere, which are still poorly understood, differ significantly from those in the outer heliosphere. As discussed in \cite{ Rankin2021ApJ_helium, Rankin2022ApJ_oxygen}, the radial component of the magnetic field predominates within 1 au, while the outer heliosphere tends to have more transverse terms; hence the transport and cross-field diffusion of particles could be different in these two regions. This might explain the higher ACR helium and oxygen radial gradient in the inner heliosphere than that beyond 1 au, regardless of the polarity of the Sun \citep{marquardt2018AA, Rankin2022ApJ_oxygen}.

Meanwhile, as the solar cycle progresses, the highly tilted HCS and increased solar activity may further disturb the large-scale IMF \citep{stone_tilt_1999}, thereby playing a role in shaping the radial gradient. The radial gradients vary over a solar cycle from solar minimum to solar maximum \citep{cummings_anomalous_1999, stone_location_2003, leske_intensity_2005} based on the measurements from Voyager in outer heliosphere. Within 1 au, ACRs are almost depleted and nearly unmeasurable during solar maximum \citep{leske_anomalous_2013,giacalone_anomalous_2022}, hence such determination has not been investigated.

\begin{figure}[!htb]
    \centering
    \resizebox{\hsize}{!}{\includegraphics{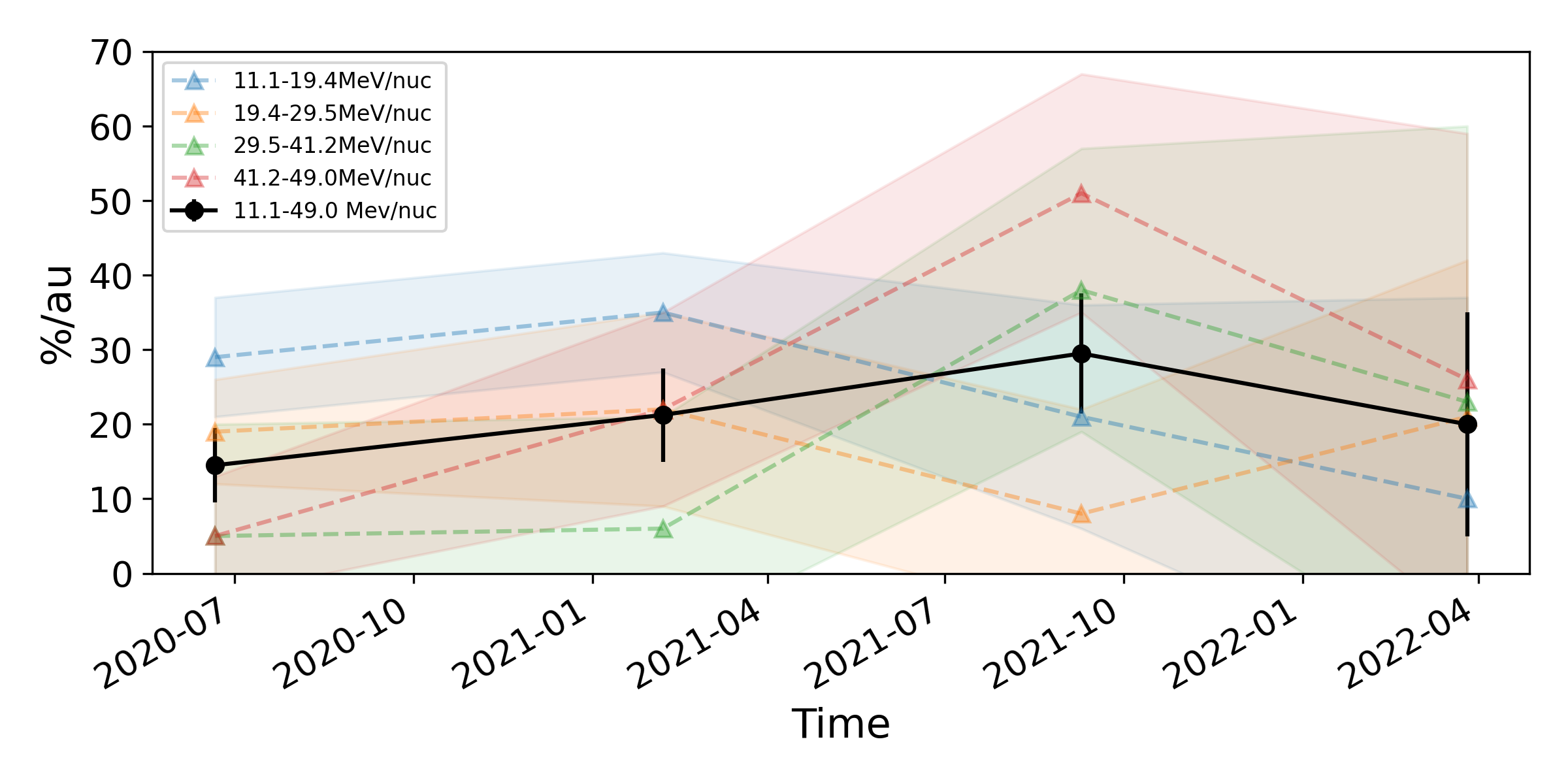}}
    \caption{Orbit-wise variations of the ACR helium radial gradients from February 2020 to July 2022 across four energies.}
    \label{fig6:radialgradient_time_variation}
\end{figure}

Therefore, based on the available measurements from the first four orbits of Solar Orbiter, we attempt to estimate the temporal variation of the radial gradients as the solar modulation progresses, and the results are displayed in Fig.~\ref{fig6:radialgradient_time_variation}. The orbiter-wise radial gradients at different energies are displayed in different colors. The shaded regions represent the relatively large uncertainties of the radial gradients due to the limited counting statistics, and the black line shows the average radial gradient across four energies. 
The averaged radial gradients for 11 - 49 MeV/nuc ACR helium increases from 15 $\pm$ 5 \%/au during the first orbit to 29.5 $\pm$ 8 \%/au in the third orbit. In the fourth orbit, the radial gradient drops to 20 $\pm$ 15 \% / au, with a considerably larger uncertainty.
It should be noted that the temporal variations in Fig.~\ref{fig6:radialgradient_time_variation} have not considered the GCR contribution to the ACR intensities, and hence, the above values are only the lower limits.
After subtracting the estimated GCR intensity, the average radial gradient rises from 23.5 $\pm$ 8\% /au at the first orbit to 81 $\pm$ 20 \%/au in the third orbit. While in the fourth orbit, the corrected radial gradient has an uncertainty larger than the radial gradient; therefore, we ignore this period.
Despite the large uncertainties, the overall trend of the mean gradient seems evident nonetheless, i.e., as the solar cycle progresses, the radial gradients of ACR helium tend to strengthen within 1 au.
\cite{cummings_radial_2009} reported that when the tilt angle of HCS is larger than 30 degrees, ACRs are unable to propagate into the inner heliosphere via the drift process. The temporal evolution revealed by our results seems consistent with this scenario and show how the radial gradient evolved with the increase of the solar activity.

Lastly, we wish to point out a dissimilarity in the ACR helium temporal profiles between HET and EPHIN. The affected data points have been excluded from our present radial gradient analysis and won't influence the current results.
Between 2020 November and 2020 December, i.e., in the 11th Solar Orbiter-perspective Carrington rotation, the helium intensity ratio of Solar Orbiter to SOHO/EPHIN is unusually higher. The intensity discrepancies of no less than 20 \% are found across all four channels, with a maximum ratio of 1.6 observed in the 40 - 50 MeV/nuc range, as the third panel of Fig.~\ref{fig:overview_helium_intensity} shows. Since the intensity offset only occur in one Carrington rotation period, one might speculate that such discrepancies arise from the presence of local magnetic structures, such as large-scale ICMEs or strong shocks that affect one location but not the other, and thereby lead to differences in intensity. Indeed, ICMEs were detected by both Solar Orbiter and WIND during the corresponding Carrington rotation, as indicated by in situ measurements and the ICMECAT catalog. But, all observations obtained during these ICME intervals have already been excluded from our analysis. Therefore, whether and how these structures affect ACR intensities still warrants further investigation.
Intriguingly, the higher ratio coincides with a local peak in the averaged sunspot number and tilt angle, as the red and blue lines in the fourth panel of Fig.~\ref{fig:overview_helium_intensity} indicate. It could be a sign that the enhanced modulations impact the two locations differently, which is not yet understood.

On 2025 February 18 Solar Orbiter finished its final gravity assist maneuver of Venus (VGAM) and gradually  move to higher latitudes to study the polar region, the last unexplored region of the Sun. This will provide a another valuable opportunity to measure the latitude gradient of energetic particles and to investigate the drift direction of the particles in the inner heliosphere.
Together with PSP orbiting the Sun within the ecliptic plane, and the newly launched Interstellar Mapping and Acceleration Probe (IMAP, \cite{mccomas_interstellar_2025}) mission at the L1 point, we may be able to further answer how the cosmic rays diffuse radially, longitudinally and latitudinally within the inner heliosphere from the equator plane to the polar region and from the sun to the outer heliosphere.

In short, our results confirm the enhanced radial gradient reported by \cite{Rankin2021ApJ_helium} at PSP within 1 au, which is larger than the previous measurements in the outer heliosphere.
Our results will serve as a constraint on the transport modeling of ACRs in the inner heliosphere.
Linking our current observations to the previous results and the present knowledge of cosmic ray can significantly advance our understanding of the role of drift and diffusion in the transport of cosmic rays and the process that shapes our heliosphere.

\begin{acknowledgements} 
Solar Orbiter is a mission of international cooperation be-tween ESA and NASA, operated by ESA. This work was supported by the German Federal Ministry for Economic Affairs and Energy and the German Space Agency (Deutsches Zentrum für Luft- und Raumfahrt, e.V., (DLR)), grant number 50OT2002.
Special thanks are given to C. Cohen and A. Cummings for multiple rounds of insightful discussion. The author also thanks Y. Wang and C. Corti for their assistance with the GCR model data. 

\end{acknowledgements}

\bibliographystyle{aa}
\bibliography{main}

\begin{appendix}
    
\onecolumn

\section{Level-2 trigger and SEP determination}

\label{Appendix_SEP_List}

The methodology for the SEP period determination basically follows the 3-sigma method, i.e., SEP periods are defined as consecutive time intervals when the flux exceeds three times the standard deviation above the background. This method is commonly used to determine the onset time of SEP events, which could be further used for velocity dispersion analysis. Additionally, we added an extra 6-hour margin on the onset and end time, assuring the SEP contamination is largely excluded. The event list remains stable when different detection thresholds (i.e., different sigma levels) are applied.

Instead of using the actual helium measurements that have lower statistics, we utilized the higher count rate products including protons of SOHO/EPHIN in the energy range below 10 MeV and the level-2 trigger rate of HET to determine the possible SEP events. Such trigger rate is a multiple detector counter, namely HET$_{\operatorname{any}(\text{A1, B1i)}}$ of Solar Orbiter/HET. 
It registers particles that can trigger any of those two SSDs, i.e., detectors A and the inner sector, which is what $i$ stands for, of detector B of sunward telescope, in coincidence without the requirements to resolve the particle species and their primary energies. This counter has a large geometry factor, hence, it has better counting statistics, leading to a better determination of the onset and the end of the SEP events, than the nominal scientific data products.
The hourly count rates of this level-2 trigger are approximately 600 to 800 during quiet time. The completed SEP periods that we removed separately from the data are given below in Table~\ref{tab:solo-sep} and Table~\ref{tab:ephin-sep}.

\begin{table*}[ht!]
    \centering
        \caption{The removed SEP periods of Solar Orbiter, determined from the  HET$_{any(a1,b1i)}$ counter.}
    \begin{tabular}{cccc}
    \hline
    No. & Start time & End time & Duration(hours) \\
    \hline
    1 & 2020-06-10 08:10:18 & 2020-06-11 10:10:18 & 26.0\\ 
    2 & 2020-07-20 18:31:24 & 2020-07-22 00:21:24 & 29.8\\ 
    3 & 2020-11-16 19:31:41 & 2020-11-18 04:01:41 & 32.5\\ 
    4 & 2020-11-20 08:51:41 & 2020-11-21 17:41:41 & 32.8\\ 
    5 & 2020-11-24 03:01:42 & 2020-11-25 10:41:42 & 31.7\\ 
    6 & 2020-11-29 01:31:42 & 2020-12-08 02:31:43 & 217.0\\ 
    7 & 2020-12-09 10:11:44 & 2020-12-12 13:51:44 & 75.7\\ 
    8 & 2021-04-17 05:32:03 & 2021-04-20 22:42:03 & 89.2\\ 
    9 & 2021-05-07 08:02:05 & 2021-05-10 21:22:06 & 85.3\\ 
   10 & 2021-05-22 09:52:08 & 2021-05-24 15:02:08 & 53.2\\ 
   11 & 2021-05-28 11:42:08 & 2021-05-30 04:12:09 & 40.5\\ 
   12 & 2021-06-22 23:02:12 & 2021-06-24 12:12:12 & 37.2\\ 
   13 & 2021-07-15 10:32:15 & 2021-07-24 02:52:16 & 208.3\\ 
   14 & 2021-08-26 06:32:21 & 2021-08-29 05:42:22 & 71.2\\ 
   15 & 2021-09-20 07:42:25 & 2021-09-22 06:02:25 & 46.3\\ 
   16 & 2021-09-27 19:02:26 & 2021-09-29 10:52:26 & 39.8\\ 
   17 & 2021-10-08 18:42:28 & 2021-10-12 16:12:28 & 93.5\\ 
   18 & 2021-10-28 03:42:31 & 2021-11-07 12:52:32 & 249.2\\ 
   19 & 2021-11-09 05:22:33 & 2021-11-12 10:52:33 & 77.5\\ 
   20 & 2021-12-04 20:52:36 & 2021-12-06 23:42:36 & 50.8\\ 
   21 & 2022-01-18 06:02:43 & 2022-01-23 20:32:43 & 134.5\\ 
   22 & 2022-01-30 16:02:45 & 2022-01-31 19:02:45 & 27.0\\ 
   23 & 2022-02-15 14:32:47 & 2022-02-26 07:42:48 & 257.2\\ 
   24 & 2022-03-02 06:42:49 & 2022-03-03 08:42:49 & 26.0\\ 
   25 & 2022-03-10 09:02:50 & 2022-03-12 11:32:51 & 50.5\\ 
   26 & 2022-03-14 05:52:51 & 2022-03-16 15:12:51 & 57.3\\ 
   27 & 2022-03-20 17:32:52 & 2022-03-24 11:22:52 & 89.8\\ 
   28 & 2022-03-28 00:52:53 & 2022-04-08 20:02:55 & 283.2\\ 
   29 & 2022-04-08 23:42:55 & 2022-04-10 12:32:55 & 36.8\\ 
   30 & 2022-04-13 01:22:55 & 2022-04-18 13:52:56 & 132.5\\ 
   31 & 2022-04-19 17:22:56 & 2022-04-26 22:34:02 & 173.2\\ 
   32 & 2022-04-27 15:02:58 & 2022-04-28 17:32:58 & 26.5\\ 
   33 & 2022-04-30 05:22:58 & 2022-05-03 11:32:58 & 78.2\\ 
   34 & 2022-05-09 20:52:59 & 2022-05-12 21:43:00 & 72.8\\ 
   35 & 2022-05-26 21:23:02 & 2022-05-28 03:13:02 & 29.8\\ 
   36 & 2022-06-07 07:43:04 & 2022-06-09 22:33:04 & 62.8\\ 
   37 & 2022-06-13 12:53:05 & 2022-06-16 07:43:05 & 66.8\\ 
   38 & 2022-07-23 10:33:10 & 2022-08-01 19:23:12 & 224.8\\
   \hline

    \end{tabular}

      \label{tab:solo-sep}
\end{table*}

\begin{table*}[ht!]
    \centering
        \caption{The SEP periods of EPHIN, determined from the 4 - 8 MeV proton time profile}
    \begin{tabular}{cccc}
    \hline
    No. & Start time & End time & Duration(hours) \\
    \hline
    1 & 2020-04-20 12:00:00 & 2020-04-21 23:00:00 & 35.0\\ 
    2 & 2020-07-17 23:00:00 & 2020-07-18 12:00:00 & 13.0\\ 
    3 & 2020-08-14 22:00:00 & 2020-08-16 06:00:00 & 32.0\\ 
    4 & 2020-10-16 03:00:00 & 2020-10-19 11:00:00 & 80.0\\ 
    5 & 2020-11-29 07:00:00 & 2020-12-18 14:00:00 & 463.0\\ 
    6 & 2021-01-06 04:00:00 & 2021-01-07 18:00:00 & 38.0\\ 
    7 & 2021-04-17 15:00:00 & 2021-04-22 21:00:00 & 126.0\\ 
    8 & 2021-05-03 11:00:00 & 2021-05-04 19:00:00 & 32.0\\ 
    9 & 2021-05-08 07:00:00 & 2021-05-15 10:00:00 & 171.0\\ 
   10 & 2021-05-22 12:00:00 & 2021-05-25 10:00:00 & 70.0\\ 
   11 & 2021-05-26 00:00:00 & 2021-05-28 00:00:00 & 48.0\\ 
   12 & 2021-05-28 12:00:00 & 2021-06-04 09:00:00 & 165.0\\ 
   13 & 2021-06-09 02:00:00 & 2021-06-14 22:00:00 & 140.0\\ 
   14 & 2021-06-25 03:00:00 & 2021-06-27 10:00:00 & 55.0\\ 
   15 & 2021-06-28 01:00:00 & 2021-06-29 23:00:00 & 46.0\\ 
   16 & 2021-07-02 18:00:00 & 2021-07-08 07:00:00 & 133.0\\ 
   17 & 2021-07-08 22:00:00 & 2021-07-31 03:00:00 & 533.0\\ 
   18 & 2021-08-26 17:00:00 & 2021-08-30 00:00:00 & 79.0\\ 
   19 & 2021-08-30 09:00:00 & 2021-09-01 08:00:00 & 47.0\\ 
   20 & 2021-09-07 15:00:00 & 2021-09-09 19:00:00 & 52.0\\ 
   21 & 2021-09-17 06:00:00 & 2021-10-02 02:00:00 & 356.0\\ 
   22 & 2021-10-09 00:00:00 & 2021-10-17 14:00:00 & 206.0\\ 
   23 & 2021-10-28 05:00:00 & 2021-11-17 11:00:00 & 486.0\\ 
   24 & 2021-12-04 01:00:00 & 2021-12-09 11:00:00 & 130.0\\ 
   25 & 2021-12-20 01:00:00 & 2021-12-22 12:00:00 & 59.0\\ 
   26 & 2021-12-23 08:00:00 & 2021-12-24 13:00:00 & 29.0\\ 
   27 & 2022-01-06 15:00:00 & 2022-01-07 06:00:00 & 15.0\\ 
   28 & 2022-01-14 02:00:00 & 2022-01-27 14:00:00 & 324.0\\ 
   29 & 2022-01-29 13:00:00 & 2022-02-04 17:00:00 & 148.0\\ 
   30 & 2022-02-06 09:00:00 & 2022-02-08 16:00:00 & 55.0\\ 
   31 & 2022-02-09 00:00:00 & 2022-02-10 00:00:00 & 24.0\\ 
   32 & 2022-02-10 21:00:00 & 2022-02-15 00:00:00 & 99.0\\ 
   33 & 2022-02-15 16:00:00 & 2022-03-08 00:00:00 & 488.0\\ 
   34 & 2022-03-08 03:00:00 & 2022-03-09 16:00:00 & 37.0\\ 
   35 & 2022-03-10 11:00:00 & 2022-04-10 10:00:00 & 743.0\\ 
   36 & 2022-04-10 19:00:00 & 2022-04-15 00:00:00 & 101.0\\ 
   37 & 2022-04-18 12:00:00 & 2022-04-26 06:00:00 & 186.0\\ 
   38 & 2022-04-27 06:00:00 & 2022-05-06 03:00:00 & 213.0\\ 
   39 & 2022-05-06 05:00:00 & 2022-05-07 10:00:00 & 29.0\\ 
   40 & 2022-05-08 17:00:00 & 2022-05-17 18:00:00 & 217.0\\ 
   41 & 2022-05-20 02:00:00 & 2022-05-25 00:00:00 & 118.0\\ 
   42 & 2022-05-25 09:00:00 & 2022-05-28 19:00:00 & 82.0\\ 
   43 & 2022-06-07 00:00:00 & 2022-06-13 00:00:00 & 144.0\\ 
   44 & 2022-06-13 03:00:00 & 2022-06-27 04:00:00 & 337.0\\ 
   45 & 2022-06-30 03:00:00 & 2022-07-07 22:00:00 & 187.0\\ 
   46 & 2022-07-09 02:00:00 & 2022-07-19 14:00:00 & 252.0\\ 
   47 & 2022-07-22 00:00:00 & 2022-07-24 00:00:00 & 48.0\\ 
   48 & 2022-07-24 10:00:00 & 2022-08-10 00:00:00 & 398.0\\
   \hline
    \end{tabular}

    \label{tab:ephin-sep}
\end{table*}

\end{appendix}

\end{document}